\documentclass[aps,prd,reprint,showpacs,amsmath,amssymb,superscriptaddress]{revtex4-1}
\usepackage{graphicx,times}
\pdfoutput=1

\begin{document}

\title{
	Sound Absorption by Subwavelength Membrane Structures:
	A Generalized Perspective
	}

\author{Min Yang}
\author{Yong Li}
\author{Chong Meng}
\author{Caixing Fu}
\affiliation{Department of Physics, Hong Kong University of Science
and Technology, Clear Water Bay, Kowloon, Hong Kong, China}
\author{Jun Mei}
\affiliation{Department of Physics, South China University of
Technology, Guangzhou 510640, China}
\author{Zhiyu Yang}
\affiliation{Department of Physics, Hong Kong University of Science
and Technology, Clear Water Bay, Kowloon, Hong Kong, China}
\author{Ping Sheng}
\email{Correspondence email: sheng@ust.hk}
\affiliation{Department of Physics, Hong Kong University of Science
and Technology, Clear Water Bay, Kowloon, Hong Kong, China}

\begin{abstract}

Decorated membrane, comprising a thin layer of elastic film with small
rigid platelets fixed on top, has been found to be an efficient
absorber of low frequency sound. In this work we consider the problem
of sound absorption from a perspective aimed at deriving upper bounds
under different scenarios, i.e., whether the sound is incident from
one side only or from both sides, and whether there is a reflecting
surface on the back side of the membrane. By considering the
negligible thickness of the membrane, usually on the order of a
fraction of one millimeter, we derive a relation showing that the sum
of the incoming sound waves' (complex) pressure amplitudes, averaged
over the area of the membrane, must be equal to that of the outgoing
waves. By using this relation, and without going to any details of the
wave solutions, it is shown that the maximum absorption achievable
from one-side incident is 50\%, while the maximum absorption with a
back reflecting surface can reach 100\%. The latter was attained by
the hybridized resonances. All the results are shown to be in
excellent agreement with the experiments. This generalized
perspective, when used together with the Green function formalism, can
be useful in gaining insights and delineating the constraints on what
are achievable in scatterings and absorption by thin film structures.

\end{abstract}
\maketitle

\section{Introduction}

A sound absorber converts the airborne acoustic energy into thermal
motions via irreversible processes. High efficiency of such processes
requires not only properly matched impedance for the absorber, but
also the dissipative ability to absorb the incident energy.
Traditional means of acoustic absorption make use of porous and
fibrous materials \cite{arenas2010recent}, gradient index materials,
or perforated panels \cite{fuchs2006micro} with tuned cavity depth
behind the panels.  They generally result in either imperfect
impedance matching to the incoming wave, or very bulky structures with
dimensions comparable to the wavelength, usually on the order of
meters for low frequency airborne sound.

Membrane-type acoustic metamaterial, consisting of decorated membrane
resonator (DMR) of various forms, has been shown to display diverse
functionalities such as efficient reflection
\cite{yang2008membrane,yang2010acoustic,naify2011transmission,ma2013low,chen2014analytical1},
enhanced transmission \cite{park2013giant}, reversed Doppler effect
\cite{lee2010reversed}, and near-field amplification
\cite{park2011amplification}. The reason for such extraordinary
behaviors can be attributed to the effective negative mass density
\cite{yang2008membrane,lee2009acoustic} and/or negative refractive
index \cite{lee2010composite,yang2013coupled} introduced by the DMR's
subwavelength resonances.  Recent works also show that the DMR can
efficiently absorb low frequency sound even with its negligible
thickness \cite{mei2012dark,chen2014analytical2}, a capability clearly
beyond what is achievable by conventional acoustic absorbers.  This
has been attributed to the high energy density of DMR's lateral
resonances.  Hybridization of different resonances can even lead to
perfect absorptions in the deep-subwavelength regime, thereby
realizing the exact time-reversed counterpart of an acoustic point
source \cite{de2002overcoming}. The latter is known to have important
applications for time-reversal wave technology
\cite{derode1995robust}. By using hybridized modes, an acoustic
metasurface can be realized with an array of such acoustic ``sinks''
with extraordinary sound absorption characteristics
\cite{ma2014acoustic}.

In this work, we present a generalized perspective, based on DMR's
special geometric characteristic, for understanding DMR's sound
absorption behaviors. It is shown that, owing to membrane's negligible
thickness, there is an equality relating the sum of mean complex
pressure amplitudes (MCPA) of the incoming waves, averaged over the
area of DMR, to that of the outgoing waves. By drawing analogy to
momentum conservation law involving two equal-mass particles, it is
easily seen that there is a component of the total incident energy,
corresponding to the center of mass motion in the two particles
collision, which is always conserved and therefore cannot be
dissipated. Hence only the energy in excess of the conserved component
is available for dissipation. This observation leads to upper limits
for DMR's absorption performance under various scenarios. In
particular, for wave incident from one side only, only half of the
incoming energy is available for dissipation, thus the absorption
percentage cannot exceed 50\% \cite{chen2014analytical2}. We verify
this conclusion by experimentally observing the sound scattering from
DMRs, and the results show excellent agreement with theoretical
predictions.  Such general considerations also show that the higher
than 50\% absorption can be achieved either by allowing waves to be
incident from both sides of the membrane, such as in the coherent
perfect absorption (CPA) scenario \cite{chong2010coherent}, or if
multiple scatterings are introduced into the system by introducing a
back reflecting surface. As an example, the acoustic metasurface with
hybrid resonances, which has been demonstrated to completely absorb
waves incident from one side, is analyzed in light of the MCPA
conservation.  The outcome, while matching the results previously
reported \cite{ma2014acoustic}, also yields additional insights.

In what follows, Section \ref{sec:surface_impedance_and_dissipation}
shows how the area-averaged response of a subwavelength membrane
structure can be reduced to a one-dimensional problem, and the related
physics, as embedded in the Green function formalism, is considerably
simplified as a result. In Section
\ref{sec:relation_involving_MCPA} we derive a general
conservation rule governing the scatterings from thin membrane
structures, and show how some conclusions can be easily derived from
this rule without having to resort to the wave equation. In Section
\ref{sec:decorated_membrane_with_semicircular_platelets} we analyze
the problem of an efficient thin membrane absorber using the
generalized perspective, and verify experimentally the derived
conservation rule. In Section
\ref{sec:perfect_absorption_by_acoustic_metasurface} the problem of a
membrane resonator coupled to a thin, sealed air cell is analyzed,
leading to the condition for attaining perfect absorption through
hybrid resonances. This is followed by a brief summary in Section
\ref{sec:concluding_remarks} that concludes the article. In the
Appendix we derive the formulas for obtaining the theoretically
relevant parameters from four-probe impedance tube data.

\section{Surface Impedance and Dissipation}
\label{sec:surface_impedance_and_dissipation}

\begin{figure}
	\includegraphics[]{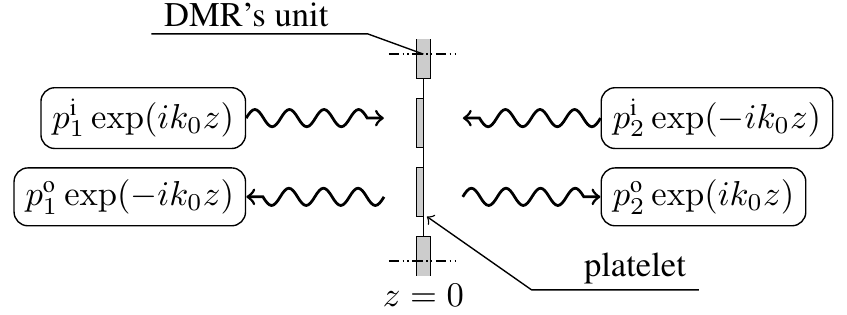}
	\caption{
		Schematic illustration of the scattering process from a DMR's
		unit. With left-incident sound wave's pressure denoted as
		$p_1^\text{i}\exp(ik_0z)$ and that from the right as
		$p_2^\text{i}\exp(-ik_0z)$, the scattering generates two
		outgoing waves $p_1^\text{o}\exp(-ik_0z)$ and
		$p_2^\text{o}\exp(ik_0z)$. Here $k_0$ is the sound wave's
		wavevector in air, and the double dot-dash lines stand for
		periodic boundary conditions.  According to the conservation
		of mean complex pressure amplitude (MCPA),
		$p_1^\text{i}+p_2^\text{i}=p_1^\text{o}+p_2^\text{o}$.
	}
	\label{fig:scatterings}
\end{figure}

A DMR unit comprises a uniformly stretched elastic membrane decorated
with relatively rigid platelets. It is important for our subsequent
considerations to note that the thickness of the membrane is
negligible ($\sim200\text{ }\mu\text{m}$), while the cross-sectional
dimension of the membrane is subwavelength in scale for the frequency
regime of interest. Without the loss of generality, we can consider it
as an inhomogeneous membrane which can vibrate and scatter the
incoming plane waves as shown in Fig.~\ref{fig:scatterings}.

Central to the understanding of DMR's acoustic behaviors is that only
the piston-like component of its averaged displacement, $\langle
W\rangle$, couples to the radiative wave modes. Here $W$ denotes the
normal displacement field of the membrane, and the angular bracket
denotes surface averaging.  However, the variance of the displacement,
$\delta W\equiv W-\langle W\rangle$, is decoupled from the radiation
modes and can be characterized as ``deaf''.  The reason of this
decoupling can be seen from the Fourier wavevectors ${\bf k}_\|$ that
delineate the lateral spatial pattern of $W$. For the $\delta W$, the
relevant ${\bf k}$ must have magnitude that satisfy the inequality
$|{\bf k}_\||>2\pi/\lambda$, where $\lambda$ is the sound wavelength
in air, owing to membrane's subwavelength cross-sectional dimension.
From the displacement continuity condition and the wave dispersion
relation, we have $(k_\|)^2+(k_\perp)^2=(2\pi/\lambda)^2$ for the
acoustic wave in air, where $k_\perp$ denotes the wavevector component
normal to the membrane. It follows that the $\delta W$ component
couples only to the evanescent waves, as its associated $k_\perp$ must
be imaginary.  In contrast, because the ${\bf k}_\|$ components for
$\langle W\rangle$ have a distribution that peaks at ${\bf k}_\|=0$,
it can couple to the radiation modes. Hence, if we restrict our
considerations to only the radiation modes, then one can treat the
problem of DMR as essentially one-dimensional in character.

The surface impedance, which characterizes the DMR's far-field
scattering properties, can be defined by using only the $\langle
W\rangle$ component, given by $Z=\langle p\rangle/\langle \dot
W\rangle$, with $p$ denoting the total sound pressure acting on the
membrane and the over-dot denoting the time derivative. We notice
that, for harmonically oscillating waves with angular frequency
$\omega=2\pi f$, the time-averaged energy flux pointing into the
membrane (which characterizes the energy dissipation rate by the
membrane) is given by $j=\int_0^{2\pi}\text{Re}(\langle p\rangle
e^{-i\omega t}) \text{Re}(\langle\dot W\rangle e^{-i\omega t})
d(\omega t)/(2\pi)$ \cite{landau1970fluid}. Since the surface
impedance's real part, $\text{Re}(Z)$, results from the velocity
component with the same phase as pressure, hence this is the component
that characterizes DMR's dissipative capability. It follows that
$j=\langle p\rangle^2\text{Re}(Z)/(2|Z|^2)$. This dissipation is
caused by the viscous damping in membrane's relative displacements. In
particular, the displacements in the ``deaf'' component $\delta W$
dominate dissipation. Therefore, there is a relation between Re$(Z)$
and $\delta W$ as shown explicitly in what follows.

The surface-averaged Green function is defined by $\langle
G\rangle=\langle W\rangle/\langle p\rangle$. Hence, for time-harmonic
motion, we have $Z=i/(\omega\langle G\rangle)$. For frequencies in the
vicinity of one of DMR's resonances, $\langle G\rangle$ is given by
\cite{yang2014homogenization}
\begin{equation}
	\langle G\rangle=\frac{|\langle
	W_n\rangle|^2}{\rho_n(\omega_n^2-\omega^2-2i\omega\beta_n)},
	\label{eq:green_function}
\end{equation}
where $\rho_n\equiv\int_\Omega\rho|W_n|^2dV$ is a parameter related to
the displacement-weighted mass density for the membrane's $n$th
eigenmode $W_n$, $\rho$ is the local mass density, $\Omega$ the volume
of the membrane, and $\omega_n$ is the relevant angular
eigenfrequency.  If we denote the viscosity coefficient of the
membrane as $\eta$ \cite{landau1970elasticity}, then the dissipation
coefficient $\beta_n$ in Eq.~\eqref{eq:green_function} is defined by
\begin{align}
	\beta_n\equiv&
	\int_\Omega\eta(\nabla W_n^\ast\cdot\nabla W_n)dV/(2\rho_n)\nonumber\\
	=&\int_\Omega\eta(\nabla\delta W_n^\ast\cdot\nabla \delta
	W_n)dV/(2\rho_n),
	\label{eq:dissipation_coefficient}
\end{align}
In Eq.~\eqref{eq:dissipation_coefficient} we have used the fact that
$\nabla\langle W_n\rangle=0$. It follows that the relevant impedance
corresponding to frequencies in the vicinity of a resonance is given
by
\begin{equation}
	Z=\frac{i}{\omega\langle G\rangle}=
	\frac{\rho_n}{|\langle W_n\rangle|^2}\left[
		2\beta_n+\frac{i}{\omega}\left(
		\omega_n^2-\omega^2
	\right)\right].
	\label{eq:impedance}
\end{equation}

Equations~\eqref{eq:dissipation_coefficient} and \eqref{eq:impedance}
indicate that, although decoupled from radiation modes, the ``deaf''
component $\delta W$ can still affect scatterings via membrane's
dissipation coefficient $\beta_n$. This can be accomplished through
structural designs.  For instance, decorating membrane with
semicircular platelets can lead to ``flapping'' modes having a large
$\delta W$ component and hence large dissipation \cite{mei2012dark}.
However, a large dissipation sometimes may also cause impedance
mismatching, thereby leading to reflection instead of absorption.
Hence ``sufficient'' dissipation that is compatible with impedance
matching is always the key to an efficient absorber.

\section{A Relation Involving the Mean Complex Pressure Amplitudes (MCPA)}
\label{sec:relation_involving_MCPA}

In this section, we will look at the problem of thin membrane
absorption through a different perspective.  As shown in
Fig.~\ref{fig:scatterings}, the two incoming waves counter-propagate
from two sides, acting on the DMR with complex pressures amplitudes
$p_1^\text{i}$ and $p_2^\text{i}$.  After scattering, they are
converted into two outgoing waves with complex pressure amplitudes
$p_1^\text{o}$ and $p_2^\text{o}$.  Here, the subscript ``1(2)''
denotes the left (right)-hand side region and the superscript ``i(o)''
stands for incoming(outgoing) waves.  Notice that the sound's energy
flux in air, $j=\int_0^{2\pi}\text{Re}(pe^{-i\omega t}) \text{Re}(\dot
We^{-i\omega t})d(\omega t)/(2\pi)=p\dot W/2=\pm p^2/(2Z_0)$, is
opposite in directions for the incident and scattering waves. Here
$Z_0$ denotes the characteristic impedance of air.  Therefore, the
air's surface-averaged normal velocities on two sides of the membrane
are given by
\begin{subequations}
\begin{align}
\label{eq:air_displacement}
	&\langle\dot W_1\rangle
	=\frac{2j_1^\text{i}}{p_1^\text{i}}+\frac{2j_1^\text{o}}{p_1^\text{o}}
	=\frac{1}{Z_0}(p_1^\text{i}-p_1^\text{o}),\\
	&\langle\dot W_2\rangle
	=\frac{2j_2^\text{o}}{p_2^\text{o}}+\frac{2j_2^\text{i}}{p_2^\text{i}}
	=\frac{1}{Z_0}(p_2^\text{o}-p_2^\text{i}),
\end{align}
\end{subequations}
respectively. Since the thickness of the membrane is negligible, we
have $\langle\dot W_1\rangle=\langle\dot W_2\rangle$, i.e., there is
no relative motion between the two sides of the membrane. That
immediately implies that the mean complex pressure amplitude (MCPA),
$\bar p$, is conserved before and after the scattering:
\begin{equation}
	\bar p=
	\frac{1}{2}\left(p_1^\text{i}+p_2^\text{i}\right)=
	\frac{1}{2}\left(p_1^\text{o}+p_2^\text{o}\right).
	\label{eq:MCPA_conservation}
\end{equation}

Conservation of the MCPA implies that the portion of incoming energy
associated with it has to be preserved from dissipation and transfered
to radiative outgoing waves by scatterings. This is an direct analogy
to the two (equal-mass) particles scattering in classical mechanics,
if we regard the $p$'s as momenta of the two particles before and
after the collision. In that analogy Eq.~\eqref{eq:MCPA_conservation}
expresses the fact that the center of mass momentum is always
conserved before and after the collision.  For waves with amplitude
$p_1^\text{i}$ from left and $p_2^\text{i}$ from right, the overall
incoming energy flux is given by
$[(p_1^\text{i})^2+(p_2^\text{i})^2]/(2Z_0)$, while the conserved MCPA
energy is simply $\bar j=\bar p^2/Z_0$. The component of the total
energy flux available for dissipation, $\tilde j$, must be the
difference between the two, given by
\begin{align}
	\tilde j=(p_1^\text{i}-p_2^\text{i})^2/(4Z_0).
	\label{eq:available_energy}
\end{align}
If we denote the energy flux being absorbed as $j_\text{ab}$, then
$j_\text{ab}\leq\tilde j$ always, and the absorption coefficient,
$A=j_\text{ab}/(\bar j+\tilde j)$, has an upper bound of $\tilde
j/(\bar j+\tilde j)$.

Some easy conclusions can be stated immediately. First, if the
incoming wave is from one side only, then $p_2^\text{i}=0$ and $\bar
p=p_1^\text{i}/2$. It follows that $\bar j=\tilde j$ and the
absorption can at most be 50\%. Second, for $A=1$, i.e., perfect
absorption, one must have $\bar j=0$, which is only possible if
$p_1^\text{i}=-p_2^\text{i}$, so that $\bar p=0$. This corresponds to
the CPA scenario \cite{chong2010coherent}.  In general, for a given
ratio $\alpha=|p_2^\text{i}|/|p_1^\text{i}|$, the maximum absorption
always occurs when $p_1^\text{i}$ and $p_2^\text{i}$ are opposite in
phase, so that they can maximally cancel each other such that
\begin{align}
	A_\text{max}&=\frac{\tilde j}{\bar j+\tilde j}
	=\frac{(p_1^\text{i})^2+(p_2^\text{i})^2+2|p_1^\text{i}p_2^\text{i}|}
	{2(p_1^\text{i})^2+2(p_2^\text{i})^2}\nonumber\\
	&=\frac{1}{2}+\frac{\alpha}{1+\alpha^2}.
	\label{eq:max_ab} 
\end{align}

It is interesting to see under what condition(s) would $\tilde j$ be
completely dissipated when the condition of $\alpha=1$ is relaxed.
That requires the total outgoing energy flux, given by
$[(p_1^\text{o})^2+(p_2^\text{o})^2]/(2Z_0)$, be equal to the
conserved MCPA energy $\bar j=\bar p^2/Z_0$. This can occur when
$p_1^\text{o}=p_2^\text{o}=\bar p$. Since the two outgoing waves are
opposite in direction, their pressure exactly cancel each other on the
membrane. Hence the net pressure on the membrane is given by $\langle
p_\text{tot}\rangle=p_1^\text{i}-p_2^\text{i}$, while displacement
continuity means that the membrane's surface-averaged normal velocity
$\langle \dot W\rangle$ must be equal to that of air, $\langle\dot
W_1\rangle$, which is defined by Eq.~\eqref{eq:air_displacement}. Thus
$\langle \dot W\rangle=\langle \dot
W_1\rangle=(p_1^\text{i}-p_1^\text{o})/Z_0
=(p_1^\text{i}-p_2^\text{i})/(2Z_0)$ since $p_1^\text{o}=\bar
p=(p_1^\text{i}+p_2^\text{i})/2$. In other words, the requirement
boils down to
\begin{equation}
	Z=\langle p_\text{tot}\rangle/\langle\dot W\rangle=2Z_0.
	\label{eq:max_ab_condition}
\end{equation}
Therefore, as long as a DMR has a purely real surface impedance twice
the impedance of air, it has the ability to absorb all the available
energy $\tilde j$ in the incoming waves.  According to
Eq.~\eqref{eq:impedance}, such impedance is realizable at the
resonances of a DMR, i.e., when $\omega=\omega_n$, with a suitably
valued $\beta_n$.

\section{Decorated Membrane with semicircular platelets}
\label{sec:decorated_membrane_with_semicircular_platelets}

\begin{figure}
	\includegraphics[]{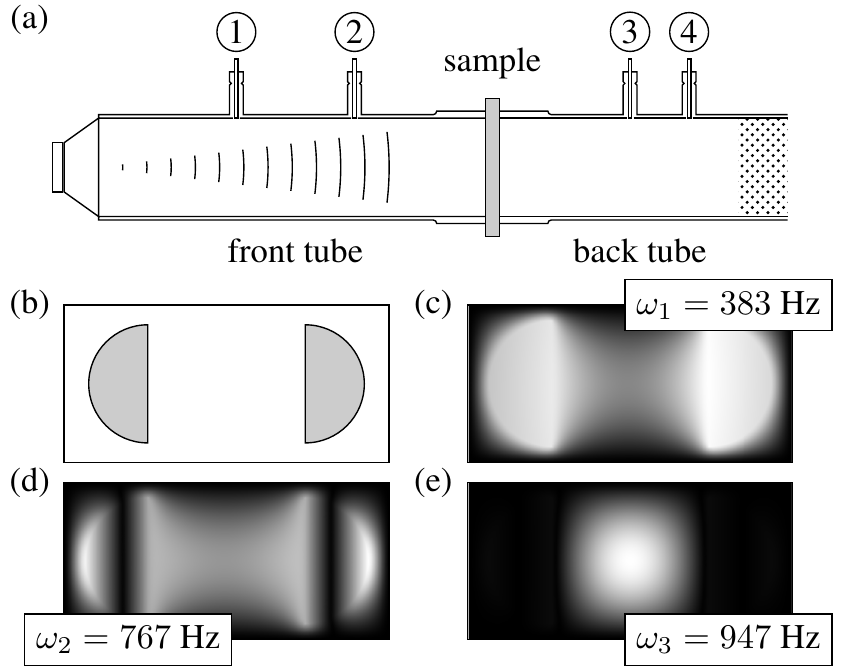}
	\caption{
		(a) Schematic illustration of the experiment apparatus. The
		front tube with a loudspeaker at one end has two sensors,
		while the back tube having the other two sensors is plugged by
		acoustic foam at another end. These four sensors are marked
		from 1 to 4 in succession. (b-e) Front views of the DMR's
		structure and its eigenmodes' profiles. Decorated by two
		opposite semicircular platelets [gray semicircles in (b)], the
		stretched membrane [white rectangle in (b)] presents three
		eigenmodes below 1 kHz whose normal displacement amplitude
		$|W|$ is plotted in gray scale as shown in (c-e).
	}
	\label{fig:sample}
\end{figure}

In this and the following Sections, we shall present two scenarios
involving large dissipation by the DMRs in which the absorption
measurements are performed by using the setup shown in
Fig.~\ref{fig:sample}(a). In the present Section we consider
absorption by DMRs with asymmetric platelets.

As the rubber membrane is usually weakly dissipative in character, we
decorate it with semicircular platelets to utilize the high
concentration of curvature energy at the edges of the platelets to
enhance the dissipation. The asymmetry of the platelets means that
there can be ``flapping'' modes [Figs.~\ref{fig:sample}(c) and (d)]
which represent a combination of normal displacement of the platelets
in conjunction with a rotational motion. At the perimeters of the
platelets, there can be very high concentration of so-called curvature
energy that is proportional to the square of the second derivative of
the normal displacement along the membrane surface directions. Since
the energy dissipation is given by the integral of the product of
energy density with the dissipation coefficient, a very high energy
density can significantly enhance the absorption of sound. Moreover,
with platelets of different weights, the size of this edge effect can
be tuned to give different $\text{Re}(Z)$.

Fig.~\ref{fig:sample}(b) shows a unit of our DMR. Four edges of the
stretched rectangle membrane are fixed on a rigid frame. On top of the
membrane, two decorated semicircular platelets are fixed to face each
other.  The membrane has a width of 15 mm, length of 30 mm, and a
thickness of 0.2 mm.  While the mass of platelets can vary in
different samples, the radius for each platelet is kept at 6 mm. 

\begin{figure}
	\includegraphics[]{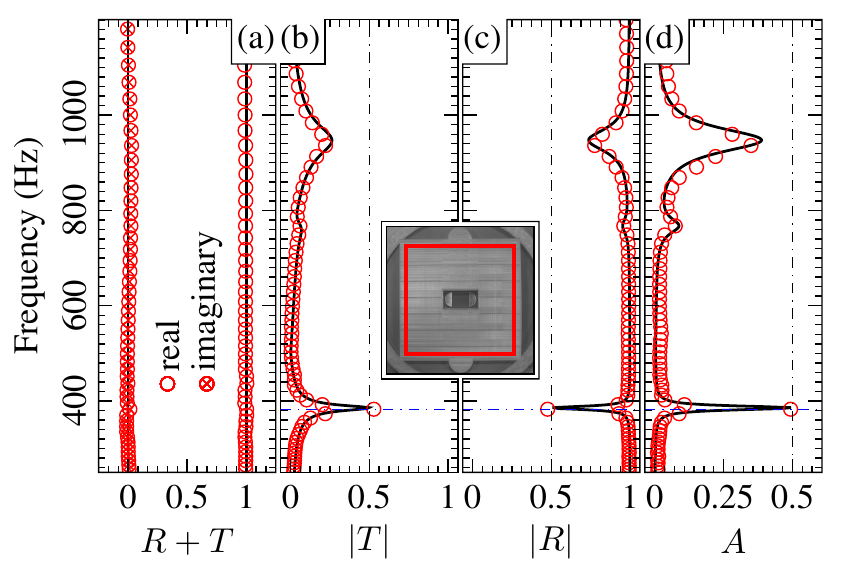}
	\caption{
		Evidence of MCPA conservation and total absorption of
		available energy under one-side incident wave. (a) Small
		discrepancies between $R+T$ and 1 ($<2.97\%$) confirm the
		conservation of MCPA. Equal amplitude, close to 0.5 for both
		$T$ in (b) and $R$ in (c) (only their magnitudes have been
		shown here for clarity) at 383 Hz, indicates all the available
		energy has been absorbed, as shown by the absorption peak of
		49.327\% at the same frequency in (d).  The red circles are
		the experimental results, the black curves are calculated from
		the eigenmodes shown in Figs.~\ref{fig:sample}(c-e). The inset
		shows the experimental sample with the red square indicating
		the area covered by the impedance tube. 
	}
	\label{fig:absorption}
\end{figure}

We test the DMR's scattering and absorption properties by mounting a
unit of the DMR on a rigid plate, and the whole sample is sandwiched
by two impedance tubes having square cross-sections (see the inset in
Fig.~\ref{fig:absorption}).  The front tube has two sensors
[Fig.~\ref{fig:sample}(a)], plus a loudspeaker at the front end to
generate the plane waves. The back tube has another two sensors, and
the tube's back end is filled by acoustic foam to eliminate
reflection.  By normalizing the pressure amplitude of all the relevant
sound waves by the incident sound pressure amplitude, the reflection
and transmission coefficients, $R$ and $T$, can be obtained from the
pressure data recorded by the four sensors (see Eq.~\ref{eq:RTcoeff}
in Appendix). 

In the present case the incident wave is from one side only, hence in
accordance to their definitions, $R=p_1^\text{o}/p_1^\text{i}$ and
$T=p_2^\text{o}/p_1^\text{i}$. The MCPA conservation in this case is
given by the simple relation $1=R+T$ (notice that this is not the
energy conservation law, which would involve the square of $R$ and
$T$). We can check it directly by comparing the measured values of
$R+T$ with 1. For a sample with identical platelets' mass of 151 mg
each, Figs.~\ref{fig:absorption}(a-c) show the measured $R+T$, $|T|$,
and $|R|$ respectively. The small discrepancy between $R+T$ and 1
($<2.97\%$) confirms the validity of the MCPA's conservation over
\emph{all} frequencies of interest.

We notice that, at the resonant frequency of 383 Hz, both $R$ and $T$
have an approximately equal value ($\simeq0.5$). It indicates an
almost total absorption for the available energy $\tilde j$ (which
equals to the MCPA energy $\bar j$ here), hence the absorption
coefficient $A$ reaches its maximum value $A_\text{max}=0.5$ [i.e.,
$\alpha=0$ in Eq.~\eqref{eq:max_ab}]. This is seen from the measured
absorption coefficient $A=1-|R|^2-|T|^2$ with a peak magnitude of
$\sim0.493$ at the same frequency as shown in
Fig.~\ref{fig:absorption}(d).

\begin{figure}
	\includegraphics[]{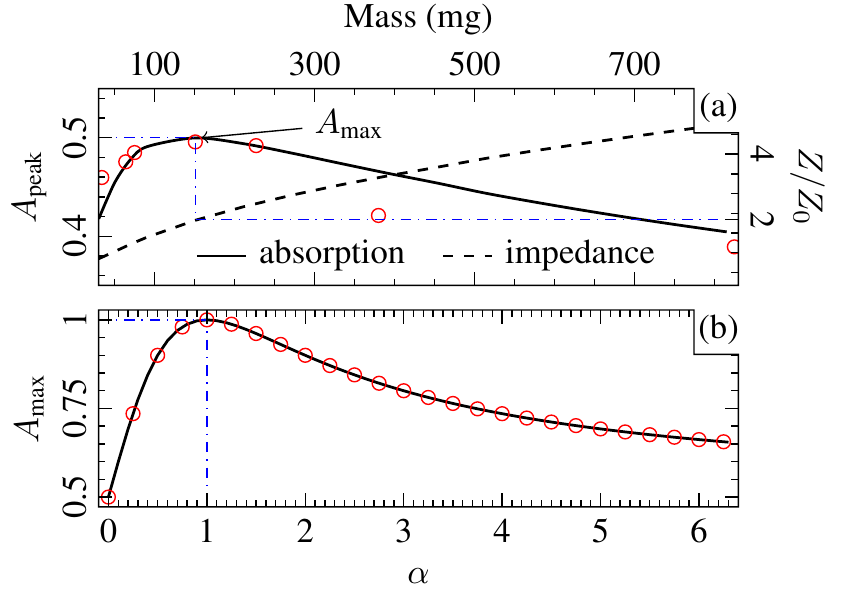}
	\caption{
		Maximum absorption for impedance-matching and its relation
		with incident waves' symmetry. (a) To absorb one-side incident
		wave, the DMR samples with differently weighted platelets are
		shown to display various peak absorption coefficients
		$A_\text{peak}$, with the maximum achieved when the relevant
		impedance satisfies the condition as expressed by
		\eqref{eq:max_ab_condition}---$Z=2Z_0$, with $Z_0$ being the
		characteristic impedance of air. The black lines are from
		numerical simulations while the red circles denote
		experimental data.  (b) Different maximum absorption
		coefficient $A_\text{max}$ for two-side incident waves for
		various values of $\alpha^2$. The numerically simulated result
		(red circles) matches Eq.~\eqref{eq:max_ab} (black curve) very
		well.
	}
	\label{fig:ab_max}
\end{figure}

The displacement profile of this (maximum absorption) resonant mode is
shown in Fig.~\ref{fig:sample}(c) [together with
Figs.~\ref{fig:sample}(d) and (e) for the other two absorption peaks
in Fig.~\ref{fig:absorption}(d)], which is numerically simulated
(based on the material parameters in
Ref.~[\onlinecite{yang2014homogenization}]) by using COMSOL
Multiphysics---a commercial finite-element solver software.  Based on
this eigenmode, the evaluated parameter values from the experimental
data are $|\langle W_1\rangle|^2/\rho_1=0.039\text{ m}^2/\text{kg}$,
$\omega_1=2\pi\times283.43$ Hz, $\beta_1=15.74$ Hz and $\eta=1.76$
Pa$\cdot$s.  The surface impedance of this DMR can therefore be
calculated by using Eq.~\eqref{eq:impedance} to yield
$Z=\text1.986Z_0$, which is very coincident with the condition as
expressed by Eq.~\eqref{eq:max_ab_condition}.  Resonances' departures
from this condition are caused by mismatching in impedance, thereby
leading to weaker absorptions.  We demonstrate this effect by 7
samples having platelets' mass varying from 34 to 825 mg, and
comparing the peak absorption coefficient $A_\text{peak}$ with the
relevant surface impedance in Fig.~\ref{fig:ab_max}(a).  Clearly, the
maximum peak absorption is attained when the condition expressed by
\eqref{eq:max_ab_condition} is met.

Through numerical simulations, we further locate the maximum
absorption coefficient $A_\text{max}$ for two-side incoming waves with
various values of $\alpha^2=|p_2^\text{i}|^2/|p_1^\text{i}|^2$, by
using the DMR with platelets' mass of 151 mg each. As shown in
Fig.~\ref{fig:ab_max}(b), almost perfect agreement between the
numerically simulated results and that predicted by
Eq.~\eqref{eq:max_ab} is seen. This confirms again the validity of the
``available energy for dissipation'' concept, and its relation to
MCPA.

\begin{figure}
	\includegraphics[]{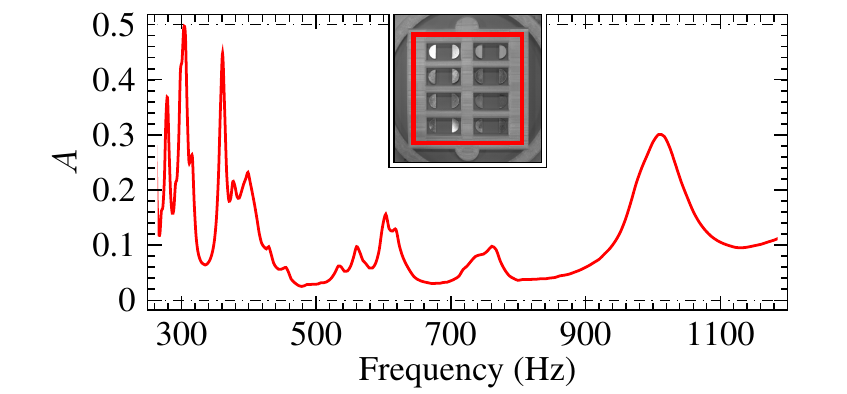}
	\caption{
		Experimentally measured absorption spectrum for a DMR panel
		with 8 units. Five in these eight units have the same
		decorated semicircular platelets weighting 225 mg, while the
		other three platelets have masses of 70, 140, and 445 mg. The
		inset shows the experimental sample, in which the red square
		indicates the area covered by the impedance tube.
	}
	\label{fig:ab_8_units}
\end{figure}

Combinations of different units in one panel can give rise to a
broader absorption spectrum. We demonstrate this by using a panel
consisting of eight different units (see the inset of
Fig.~\ref{fig:ab_8_units}).  Five of them have decorated platelets
with mass of 225 mg each, and the other three have platelets with
masses of 70, 140, and 445 mg. The experimentally measured absorption
coefficient $A$ for this panel is shown in Fig.~\ref{fig:ab_8_units}.
A relatively broad absorption spectrum [compared to that shown in
Fig.~\ref{fig:absorption}(d)] is seen due to the merging of multiple
absorption peaks.  We would like to note here that none of the
absorption spectra in Fig.~\ref{fig:absorption}(d) and
Fig.~\ref{fig:ab_8_units} exceed 50\%. In this context we would like
to note that in Ref.~[\onlinecite{mei2012dark}], maximum absorption
over 70\% has been reported for a one-side incident wave
configuration. The reason for this discrepancy is simply the
absorption area.  The DMR panel in Ref.~[\onlinecite{mei2012dark}] has
a larger area (about 2.26 times larger) than the incoming wave front,
and part of the sample extended outside of the impedance tube. Here,
the tested samples are always contained within the impedance tube, and
have exactly the same cross-section [see the insets in
Fig.~\ref{fig:absorption}(d) and Fig.~\ref{fig:ab_8_units}].

\section{Perfect Absorption by Acoustic Metasurface}
\label{sec:perfect_absorption_by_acoustic_metasurface}

\begin{figure}
	\includegraphics[]{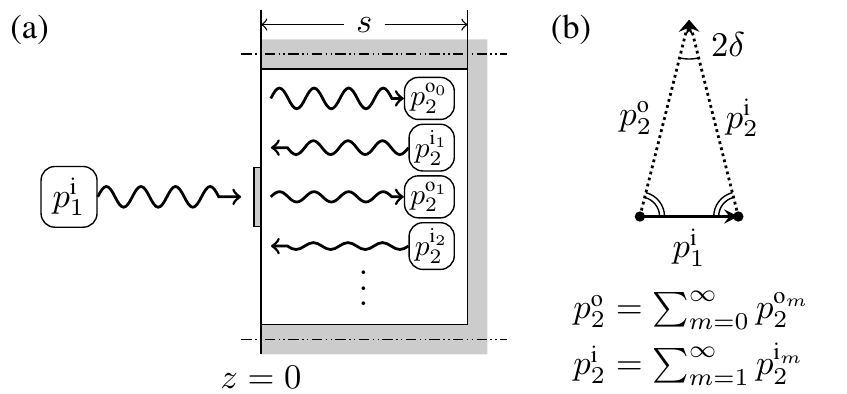}
	\caption{
		Schematic illustration of the multiply-scattered waves that
		can lead to perfect absorption on a DMR, with a reflecting
		wall placed a small distance $s$ behind it.  (a) When the
		incoming wave energy has been totally absorbed, there should
		be no reflections, i.e., all the scatterings on the incident
		side cancel each other
		$p_1^\text{o}\exp(-ik_0z)=\sum_{m=0}^\infty
		p_1^{\text{o}_m}\exp(-ik_0z)=0$.  Incoming wave
		$p_1^\text{i}\exp(ik_0z)$ generates only two net
		counter-propagating wave components
		$p_2^\text{o}\exp(ik_0z)=\sum_{m=0}^\infty
		p_2^{\text{o}_m}\exp(ik_0z)$ and
		$p_2^\text{i}\exp(-ik_0z)=\sum_{m=1}^\infty
		p_2^{\text{i}_m}\exp(-ik_0z)$, in the air layer between the
		DMR (at $z=0$) and the reflecting wall.  Here, the double
		dot-dash lines stand for the location at which the periodic
		boundary condition is applied. (b) Due to the MCPA
		conservation, the three complex pressure amplitudes must form
		an isosceles triangle whose two equal sides are slanted toward
		each other with an angle given by $2\delta=2k_0s$.
	}
	\label{fig:tot_ab}
\end{figure}

Although the MCPA energy in the incoming waves has to be conserved in
a single scattering event, its dissipation is still possible through
multiple scatterings. The outgoing MCPA energy after each scattering
can serve as the incoming energy again for the subsequent scatterings.
An example is the acoustic metasurface reported in
Ref.~[\onlinecite{ma2014acoustic}], in which a reflecting wall is
placed behind a DMR, separated by a distance $s$ that is deeply
subwavelength (2 orders smaller than the relevant wavelength in air),
thereby creating multiple reflections between the membrane and the
reflecting wall [see in Fig.~\ref{fig:tot_ab}(a)]. As a result, the
absorption coefficient for one-side incident wave can reach nearly
100\%.  In this section, we analyze this multiple-scattering process
and the perfect absorption condition with the aid of MCPA
conservation.

In the air layer between the membrane and the reflecting surface, the
outgoing wave after the $m$th scattering from the membrane,
$p_2^{\text{o}_m}$, is reflected by the wall, and becomes the incoming
wave, $p_2^{\text{i}_{m+1}}$, for the $(m+1)$th scattering. These
multiple scattered waves can be superposed to form two
counter-propagating waves: $p_2^\text{i}=\sum_{m=1}^\infty
p_2^{\text{i}_m}$ and $p_2^\text{o}=\sum_{m=0}^\infty
p_2^{\text{o}_m}$.  Because the reflecting wall presents a velocity
node, these two wave components in front of the reflecting wall must
be equal in magnitude but different in phase by $2\delta=2k_0s$ at the
position of membrane, i.e.,
\begin{equation*}
	|p_2^\text{i}|=|p_2^\text{o}|,\quad
	p_2^\text{i}/p_2^\text{o}=\exp(2i\delta).
\end{equation*}
Here $k_0$ denotes the sound's wavevector in air, and $s$ is
the air layer's thickness.

\begin{figure}
	\includegraphics[]{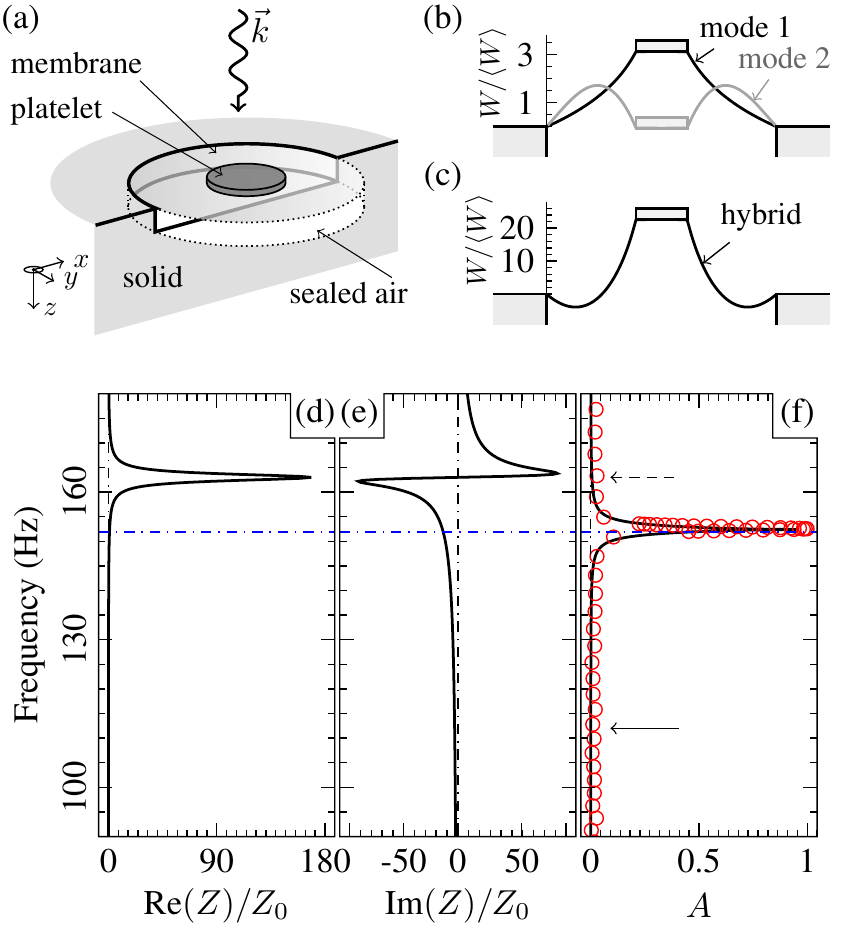}
	\caption{
		The behavior of a perfect absorber for one-side incident wave,
		denoted the hybrid resonance metasurface.  (a) Schematic
		illustration of the unit cell's component and geometry. (b)
		Schematic cross-sectional illustration of the two lowest
		frequency eigenmodes of the DMR, plotted in reference to the
		same phase of incident wave, with $W$ being the normal
		displacement of membrane, normalized to its surface average
		component $\langle W\rangle$. (c) Displacement profile of the
		DMR's hybrid mode at 152 Hz, which is clearly a superposition
		of the two in (b), but with an almost an order of magnitude
		larger amplitude.  However, such large amplitude oscillations
		are only coupled to the non-radiative evanescent waves, as
		shown in Section \ref{sec:surface_impedance_and_dissipation}.
		(d,e) As a function of frequency, the DMR's surface impedance
		$Z$, normalized to the impedance of air, $Z_0$, is seen to
		satisfy the total absorption requirement given by
		Eq.~\eqref{eq:tot_ab_condition} at the hybrid resonance
		frequency, which is indicated by the blue dot-dashed line.
		This is confirmed experimentally measuring the absorption
		coefficient, shown in (f). A sharp absorption peak, reaching
		0.994, is seen. The solid and dashed arrows indicate the first
		eigenmode and the anti-resonance frequencies of the DMR.
		Experimental results are shown in red circles, and black
		curves denote the simulation results.
	}
	\label{fig:hybrid}
\end{figure}

As the conservation of MCPA is valid for membrane's each scattering,
it also holds for their superpositions---the total wave field on the
membrane.  Therefore,
$(p_1^\text{i}+p_2^\text{i})/2=(p_1^\text{o}+p_2^\text{o})/2$, where
$p_1^\text{o}=\sum_{m=0}^\infty p_1^{\text{o}_m}$ is also a
superposition for outgoing waves on the left-hand side after each
scattering.  Assuming the incoming energy is completely absorbed by
the DMR, there will be no reflection on the incident side, thus
$p_1^\text{o}=0$. Thus the conservation of MCPA,
$p_1^\text{i}=p_2^\text{o}-p_2^\text{i}$, may be represented by a
phasor-diagram, shown in Fig.~\ref{fig:tot_ab}(b), which is an
isosceles triangle with $p_2^\text{i}$ and $p_2^\text{o}$ being the
two equal sides and $p_1^\text{i}$ being the base. It is clear from
the diagram that $p_2^\text{i}+p_2^\text{o}$, which is the right-hand
side pressure on the DMR, differs from the left-hand side pressure,
$p_1^\text{i}$, by a phase of $\pi/2$. That is,
$p_2^\text{i}+p_2^\text{o}$ is perpendicular to $p_1^\text{i}$ in the
phasor-diagram, so that
$p_2^\text{i}+p_2^\text{o}=ip_1^\text{i}\cot\delta$.  Since the
surface-averaged velocity is given by $\langle\dot
W\rangle=\langle\dot W_1\rangle=p_1^\text{i}/Z_0$, the total pressure
applied to the DMR, $\langle
p_\text{tot}\rangle=p_1^\text{i}-(p_2^\text{i}+p_2^\text{o})$, is
given by $\langle p_\text{tot}\rangle=p_1^\text{i}(1-i\cot\delta)$. It
follows that the impedance condition for achieving total absorption is
given by
\begin{equation}
	Z=\langle p_\text{tot}\rangle/\langle\dot W\rangle
	=Z_0(1-i\cot\delta).
	\label{eq:tot_ab_condition}
\end{equation}
For $\delta=\pi/2$, we have $Z=Z_0$, i.e., impedance matching. Such
condition occurs at the sealed cell's ``drum'' resonances when
$s=(N/2+1/4)\lambda$, with $N$ being an integer. However, for an air
layer thinner than a quarter-wavelength, $\delta\to0$, and the
required imaginary part of $Z$ approaches $-i\infty$. Such a large
imaginary part of the DMR impedance is necessary in order to cancel
the impedance of the reflecting back surface that is close by. Since
the impedance of the whole structure is the addition of the two serial
impedances of the DMR and the sealed cell behind it, the net impedance
of the structure is still $Z=Z_0$, i.e., impedance matched to that of
air \cite{ma2014acoustic}. In what follows, we will show that by
utilizing the \emph{hybrid mode} of the DMR, such large imaginary
impedance of the DMR can indeed be realized, while maintaining the
real part to be $Z_0$. The reason behind the large imaginary impedance
is due to the existence of DMR's anti-resonance condition, realized at
frequency $\tilde\omega$ that is in-between the two low-lying
resonances, at which the dynamic mass density of the DMR can display a
resonance-like dispersion \cite{yang2008membrane}.

In order to see the emergence of the hybrid mode, here for simplicity
we consider only two relevant eigenmodes $W_1$ and $W_2$ [as shown in
Fig.~\ref{fig:hybrid}(b)], in contrast to
Eq.~\eqref{eq:green_function}, which is valid only for frequencies
close to one of the DMR resonances. The surface-averaged Green
function of the DMR is given by \cite{yang2014homogenization}
\begin{equation}
	\langle G\rangle=\sum_{n=1}^2
	\frac{|\langle W_n\rangle|^2}{\rho_n(\omega_n^2-\omega^2)}+
	2i\beta\sum_{n=1}^2
	\frac{|\langle W_n\rangle|^2\omega}{\rho_n(\omega_n^2-\omega^2)^2}
	\label{eq:green_function_2}
\end{equation}
where $\beta$, the averaged coefficient of $\beta_1$ and $\beta_2$, is
taken to be a small quantity so that the dimensionless
$\beta/\omega\ll1$. We notice that in the vicinity of the DMR's
anti-resonance frequency $\tilde\omega$, at which $\text{Re}(\langle
G\rangle)=0$ (i.e., $Z\to\infty$), this surface-averaged Green
function behaves as $\langle G\rangle\simeq2\Xi(i\beta-\Delta\omega)$,
where
\begin{equation*}
	\Xi\equiv\sum_{n=1}^2
	\frac{|\langle W_n\rangle|^2\tilde\omega}
	{\rho_n(\omega_n^2-\tilde\omega^2)^2},\text{ and }
	\Delta\omega\equiv\tilde\omega-\omega.
\end{equation*}
From the inverse relationship between the Green function and impedance
as describe previously, the surface impedance in the vicinity of DMR's
anti-resonance is given by
\begin{equation}
	Z=\frac{1}{2\tilde\omega\Xi}\frac{1}{\beta^2+\Delta\omega^2}
	(\beta-i\Delta\omega).
	\label{eq:impedance_2}
\end{equation}
Equating Eqs.~\eqref{eq:tot_ab_condition} and \eqref{eq:impedance_2}
yields two equations for the condition of total absorption:
\begin{subequations}
	\label{eq:hybrid_ab_condition}
\begin{align}
	&\frac{1}{2\tilde\omega\Xi}\frac{\beta}{\beta^2+\Delta\omega^2}=Z_0,\\
	&\frac{1}{2\tilde\omega\Xi}\frac{\Delta\omega}{\beta^2+\Delta\omega^2}
	=Z_0\cot\delta.
\end{align}
\end{subequations}
This total absorption condition is robust since experimentally we have
two parameters, $\beta$ and $\Delta\omega$, which can be adjusted to
satisfy Eq.~\eqref{eq:hybrid_ab_condition}. Notice that this total
absorption mode of the membrane is neither of its two original DMR
resonances, but a hybridization of them [as shown in
Fig.~\ref{fig:hybrid}(c)] caused by the back reflecting surface that
is located in the near-field region of the DMR. Since the dissipation
coefficient is required to be very small, hence to completely
dissipate the incoming acoustic energy, the amplitude of this hybrid
mode is always significantly larger than the original membrane's
resonances for the same incident wave amplitude. This is clear from
the comparison between Figs.~\ref{fig:hybrid}(b) and (c). However, it
is to be especially noted that in spite of the large maximum amplitude
of the hybrid mode, the surface-averaged normal displacement is small
and matches that of air.

Experiments reported in Ref.~[\onlinecite{ma2014acoustic}] have
corroborated the existence and condition for this hybrid resonance
total absorption. By placing an aluminum-reflecting wall behind a DMR,
which is a 90 mm-wide, tensioned circular elastic membrane decorated
by a 800 mg platelet with a radius of 10 mm
[Fig.~\ref{fig:hybrid}(a)], an absorption peak with $A>0.99$ was
observed at 152 Hz [Fig.~\ref{fig:hybrid}(f)]. By using a value of
$s=21.7$ mm, which is inferred from the experimental data of $s=17$ mm
of SF6 gas, as the effect of two thin gas layers is identical if their
thicknesses are scaled linearly with their adiabatic index (SF6 has an
adiabatic index of 1.098, compared to 1.4 for air)
\cite{ma2014acoustic}.  Based on the two relevant eigenmodes shown in
Fig.~\ref{fig:hybrid}(b), the evaluated
$\Xi=1.268\times10^{-9}\text{m}^2\text{s}^3/\text{kg}$,
$\tilde\omega=2\pi\times162.3$ Hz, and $\beta=5.14$ Hz.  Equation
\eqref{eq:impedance_2} gives the surface impedance of the DMR as a
function of frequency, shown in Figs.~\ref{fig:hybrid}(d) and
\ref{fig:hybrid}(e).  At the total absorption frequency, we have
$Z=1.001-12.961i$, which is noted to match the value required,
$Z_0(1-i\cot(k_0s))=1-16.342i$, reasonably well.

\section{Concluding Remarks}
\label{sec:concluding_remarks}

We have proposed a generalized perspective for the scattering and
absorption behaviors of thin membrane structures in the subwavelength
regime. As a result of DMR's negligible thickness, the conservation of
MCPA for the incoming and outgoing waves can be derived. By analogy to
the momentum conservation law involving two equal-mass particles, it
is easy to see that within one scattering, waves incident from one
side can only be absorbed by a maximum of 50\%. Higher absorption
requires waves incident from both sides of the DMR. Based on the Green
function formalism, we determined the value of maximal absorption
under different scenarios. In particular, maximal absorption is
realized when the DMR has a surface impedance twice that of air. These
conclusions are examined experimentally. In all cases good agreement
with the theory is obtained.

For one-side incident, absorption higher than 50\% can be achieved by
introducing multiple scatterings, e.g., by placing a reflecting
surface behind the DMR. Analysis by using the MCPA conservation law
shows that near-perfect absorption can be achieved through hybrid
resonances, in excellent agreement with the experimental results
reported recently \cite{ma2014acoustic}; while at the same time it
also yields the general condition [Eq~\eqref{eq:tot_ab_condition}] for
achieving such characteristic.

Due to the similarity between the acoustic and electromagnetic
waves, there can clearly be analogous considerations for the latter,
e.g., in laser plasma absorption
\cite{godwin1972optical,freidberg1972resonant,kindel1975surface}.
Hence similar results are expected for electromagnetic wave
scatterings from thin film structures. In particular, similarity may
extend to perfect absorption by electromagnetic hybrid resonances.

\begin{acknowledgments}
	J.M. and P.S. wish to thank Guoliang Huang for helpful
	discussions.  This work is supported by Hong Kong RGC Grants
	AoE/P-02/12.  JM is supported by the National Natural Science
	Foundation of China (Grant No.  11274120), and the Fundamental
	Research Funds for the Central Universities (Grant No.
	2014ZG0032).
\end{acknowledgments}

\appendix
\section{Transmission/Reflection Retrieval Method}
\label{sec:t/r_retrieval_method}

As scalar waves, airborne sound can propagate in a sub-wavelength
waveguide without a cut-off frequency. In our experiments, the
geometrical size of the apparatus (viz., the width of the square
waveguide)  [cf. Fig.~\ref{fig:sample} (a)] is smaller than the
measured wavelength, so that only plane waves can propagate in both
the front and back tubes \cite{morse01}.  The total pressure fields in
the two (front and back) impedance tubes may be expressed as the sum
of forward and backward waves propagating along the $z$ direction:
\begin{subequations}
\label{eq:Total_pressure}
	\begin{align}
	&p_1=p_1^\text{i}{e^{i{k_0}z}} + p_1^\text{o}{e^{ - i{k_0}z}},\\
	&p_2=p_2^\text{i}{e^{-i{k_0}z}} + p_2^\text{o}{e^{i{k_0}z}}.
	\end{align}
\end{subequations}
Here the subscripts ``1" and ``2" refer to the front and back tubes
and the superscripts ``i" and ``o" represent the incoming and outgoing
waves, respectively.  To retrieve the transmission and reflection
coefficients, the total pressure fields should be exactly expressed,
by using the four experimentally measured parameters: $p_1^\text{i},
p_1^\text{o}, p_2^\text{i}$ and $p_2^\text{o}$.  Four sensors are used
to determine these parameters.  Two sensors labeled ``1" and ``2" are
placed in the front tube at $z_1=-339.5$ mm and $z_2=-239.5$ mm, and
the other two sensors labeled ``3" and ``4" are placed in the back
tube at $z_3=193.0$ mm and $z_4=393.0$ mm [cf. Fig.~\ref{fig:sample}
(a)].  According to Eq.~\eqref{eq:Total_pressure}, the pressure values
at the positions of the four sensors are 
\begin{subequations}
\label{eq:pressure_sensors}
	\begin{align}
	&p(z_1)=p_1^\text{i}{e^{i{k_0}z_1}} + p_1^\text{o}{e^{-i{k_0}z_1}},\\
	&p(z_2)=p_1^\text{i}{e^{i{k_0}z_2}} + p_1^\text{o}{e^{-i{k_0}z_2}},\\
	&p(z_3)=p_2^\text{i}{e^{-i{k_0}z_3}} + p_2^\text{o}{e^{i{k_0}z_3}},\\
	&p(z_4)=p_2^\text{i}{e^{-i{k_0}z_4}} + p_2^\text{o}{e^{i{k_0}z_4}}.
	\end{align}
\end{subequations}
By solving Eq.~\eqref{eq:pressure_sensors}, we obtain
\begin{subequations}
\label{eq:parameters}
	\begin{align}
	&p_1^\text{i}=\frac{p(z_1) e^{i{k_0}z_1}-p(z_2) e^{i{k_0}z_2}}{e^{2ik_0 z_1}-e^{2ik_0 z_2}},\\
	&p_1^\text{o}=-\frac{p(z_1) e^{i{k_0}z_2}-p(z_2) e^{i{k_0}z_1}}{e^{2ik_0 z_1}-e^{2ik_0 z_2}}e^{ik_0(z_1+z_2)} ,\\
	&p_2^\text{i}=-\frac{p(z_3) e^{i{k_0}z_4}-p(z_4) e^{i{k_0}z_3}}{e^{2ik_0 z_3}-e^{2ik_0 z_4}}e^{ik_0(z_3+z_4)} ,\\
	&p_2^\text{o}=\frac{p(z_3) e^{i{k_0}z_3}-p(z_4) e^{i{k_0}z_4}}{e^{2ik_0 z_3}-e^{2ik_0 z_4}},	
	\end{align}
\end{subequations}
Here $p(z_j)$ is the pressure measured by each sensor
labeled as ``$j=1\sim4$" in the subscripts.

The scattering matrix $S(k_0)$ describing the relationship between the
incoming and outgoing waves can be expressed as 
\begin{equation}
	\label{eq:Smatrix}
	\left( {\begin{array}{*{20}{c}}
		{p_2^\text{o}}\\
		{p_1^\text{o}}
	\end{array}} \right) = S(k_0)\left( 
	{\begin{array}{*{20}{c}}
		{p_1^\text{i}}\\
		{p_2^\text{i}}
	\end{array}} \right),
	\quad S(k_0) = \left( 
	{\begin{array}{*{20}{c}}
		T&R\\
		R&T
	\end{array}} \right).
\end{equation}
It should be noted that, due to the symmetry of the sample in our
system, the reflection and transmission coefficients $R$ and $T$ are
identical if the sample is tuned around 180 degrees.  As a result, the
reflection and transmission coefficients can be retrieved as
\begin{subequations}
\label{eq:RTcoeff}
\begin{align}
	&R=\frac{p_1^\text{i} p_1^\text{o} - p_2^\text{i}
	p_2^\text{o}}{p_1^\text{i} p_1^\text{i} - p_2^\text{i}
	p_2^\text{i}},\\
	&T=\frac{p_1^\text{i} p_2^\text{o} - p_1^\text{o}
	p_2^\text{i}}{p_1^\text{i} p_1^\text{i} - p_2^\text{i} p_2^\text{i}}.
\end{align}
\end{subequations}
Here the four wave amplitude $p_{1}^\text{i}$, $p_{1}^\text{o}$,
$p_{2}^\text{i}$ and $p_{2}^\text{o}$ are determined from
Eq.~\eqref{eq:parameters}.

\bibliography{main}

\end{document}